\newcommand{\diracslash}[1]{#1\llap{/\kern2pt}}

\newcommand{\be}{\begin{equation}}
\newcommand{\ee}{\end{equation}}
\newcommand{\bea}{\begin{eqnarray}}
\newcommand{\eea}{\end{eqnarray}}
\newcommand{\ba}[1]{\begin{array}{#1}}
\newcommand{\ea}{\end{array}}

\newcommand{\bt}{\begin{tabular}}
\newcommand{\et}{\end{tabular}}

\newcommand{\beas}{\begin{eqnarray*}}
\newcommand{\eeas}{\end{eqnarray*}}

\documentclass[preprint,prd,aps,floats,nofootinbib,floatfix]{revtex4}
\usepackage{graphicx}
\begin{document}

\title{Light vector meson masses in strange hadronic matter\\
-- a QCD sum rule approach}
\author{Amruta Mishra}
\email{amruta@physics.iitd.ac.in}
\affiliation{Department of Physics, Indian Institute of Technology, Delhi,
Hauz Khas, New Delhi -- 110 016, India}
\begin{abstract}
We study the properties of the light vector mesons ($\rho$, $\omega$
and $\phi$) in strange hadronic matter using the QCD sum rule approach.
The in-medium masses of the vector mesons are calculated from the 
modifications of the light quark condensates and the gluon condensates 
in the hadronic medium. The light quark condensates in the hadronic matter
are obtained from the values of the non strange and strange scalar fields 
of the explicit chiral symmetry breaking term in a chiral SU(3) model. 
The in-medium gluon condensate is calculated through the medium 
modification of a scalar dilaton field, which is introduced 
into the chiral SU(3) model to simulate the scale symmetry 
breaking of QCD. The mass of the $\omega$ meson is observed 
to have initially a drop with increase in density and then 
a rise due to the scattering with the baryons. The mass of 
the $\rho$ meson is seen to drop with density due to decrease 
of the light quark condensates in the medium. The effects of 
isospin asymmetry and strangeness of the medium on the masses 
of the vector mesons are also studied in the present work. 
The $\phi$ meson is observed to have marginal drop in its mass 
in the nuclear medium. However, the strangeness of the medium 
is seen to lead to an appreciable increase in its mass arising
due to scattering with the hyperons.

\end{abstract}
\maketitle

\def\bfm#1{\mbox{\boldmath $#1$}}

\section{Introduction}
The study of the properties of hadrons in hot and dense matter is an 
important topic of research in strong interaction physics.
The changes in the hadron properties in the medium affect
the experimental observables from the hot and/or
dense matter produced in the heavy ion collision experiments. 
The medium modifications of the properties of the light vector mesons 
\cite{rapp} can affect the low mass dilepton spectra, the 
properties of the kaons and antikaons can show 
in their production as well as collective flow. The modifications 
of the properties of the charm mesons, $D$ and $\bar D$ as well as 
the charmonium states can modify the yield of the open charm meson 
as well as charmonium states in the high energy nuclear collision
experiments.

In the present work, we study the medium modification of the masses 
of the light vector mesons ($\rho$, $\omega$ and $\phi$)
in the strange hadronic matter due to the interaction with the 
light quark condensates
and the gluon condensates using the QCD sum rule approach
\cite{hatlee,hatlee2,zschocke,klinglnpa,kwonprc2008,Abhee}. 
The light quark condensates are calculated from the expectation values
of the non-strange and strange scalar fields of the explicit chiral
symmetry breaking term in a chiral SU(3) model \cite{kristof1,papa}.
The gluon condensate in the hadronic medium is obtained from the medium 
modification of a scalar dilaton field introduced within the chiral 
SU(3) model
through a scale symmetry breaking term in the Lagrangian density
leading to the QCD trace anomaly. The chiral SU(3) model has been
used to describe the hadronic properties in the vacuum as well as 
in nuclear matter \cite{kristof1}, finite nuclei  \cite{papa} 
and the bulk properties of (proto) neutron stars \cite{nstar}.
The vector mesons have also been studied within the model \cite{vecm},
arising due to their interaction with the nucleons in the medium.
The model has been used to study the medium modifications 
of kaons and antikaons in isospin asymmetric nuclear matter in 
\cite{isoamss} and in hyperonic matter in \cite{isoamss2}. The chiral 
effective model has also been generalized to SU(4) to derive the interactions
of the charm mesons with the light hadrons to study the $D$ mesons 
in asymmetric nuclear matter at zero temperature \cite{amarind} 
and in the symmetric and asymmetric nuclear (hyperonic) matter 
at finite temperatures in Ref.\cite{amdmeson} and 
Ref. \cite{amarvind,amarvindhyp}. In the present investigation,
we study the light vector mesons using QCD sum rule approach
due to their interaction with the quark and gluon condensates
in the strange hadronic medium. These in-medium condensates are calculated 
in a chiral SU(3) model, from the explicit chiral symmetry breaking term and 
the scale breaking term of the Lagrangian density of the effective hadronic 
model.

The outline of the paper is as follows : In section II, we give a brief 
introduction of the chiral $SU(3)$ model used to calculate the quark and
gluon condensates in the hadronic medium. In the present work, 
the in-medium condensates as calculated in the chiral SU(3) model,
are taken as inputs for studying the in-medium masses of the light vector 
mesons using the QCD sum rule approach. The medium modifications 
of the quark and gluon condensates arise from the medium modification 
of the scalar fields of the explicit symmetry breaking term and of the 
scalar dilaton field introduced in the hadronic model to incorporate 
broken scale invariance of QCD. In section III, we present the results 
for the medium modifications of the light vector mesons using 
a QCD sum rule approach. In section IV, we summarize the findings 
of the present investigation and compare with the existing results 
in the literature for the in-medium properties of the light vector mesons.

\section{The hadronic chiral $SU(3) \times SU(3)$ model }
We use an effective chiral $SU(3)$ model for the present investigation 
\cite{papa}. The model is based on the nonlinear realization of chiral 
symmetry \cite{weinberg, coleman, bardeen} and broken scale invariance 
\cite{papa,kristof1,vecm}. This model has been used successfully to 
describe nuclear matter, finite nuclei, hypernuclei and neutron stars. 
The effective hadronic chiral Lagrangian density contains the following terms
\begin{equation}
{\cal L} = {\cal L}_{kin}+\sum_{W=X,Y,V,A,u} {\cal L}_{BW} + 
{\cal L}_{vec} + {\cal L}_{0} + {\cal L}_{SB}
\label{genlag}
\end{equation}
In Eq. (\ref{genlag}), ${\cal L}_{kin}$ is kinetic energy term, 
${\cal L}_{BW}$ is the baryon-meson interaction term in which the 
baryon-spin-0 meson interaction term generates the vacuum baryon masses. 
${\cal L}_{vec}$  describes the dynamical mass generation of the vector 
mesons via couplings to the scalar mesons and contain additionally 
quartic self-interactions of the vector fields. ${\cal L}_{0}$ contains 
the meson-meson interaction terms inducing the spontaneous breaking of 
chiral symmetry as well as a scale invariance breaking logarithmic 
potential. ${\cal L}_{SB}$ describes the explicit chiral symmetry breaking. 

To study the in-medium hadron properties using the chiral SU(3) model, 
we use the mean  field approximation,
where all the meson fields are treated as classical fields. 
In this approximation, only the scalar and the vector fields 
contribute to the baryon-meson interaction, ${\cal L}_{BW}$
since for all the other mesons, the expectation values are zero.
The baryon-scalar meson coupling constants are fitted from the 
vacuum masses of the baryons. The parameters in the model \cite{papa,isoamss} 
are chosen so as to decouple the strange vector field 
$\phi_{\mu}\sim\bar{s}\gamma_{\mu}s$ from the nucleon.

The concept of broken scale invariance leading to the trace anomaly 
in QCD, $\theta_{\mu}^{\mu} = \frac{\beta_{QCD}}{2g} 
{G^a}_{\mu\nu} G^{\mu\nu a}$, where $G_{\mu\nu}^{a} $ is the 
gluon field strength tensor of QCD, is simulated in the effective 
Lagrangian at tree level through the introduction of 
the scale breaking terms \cite{sche1,ellis}
\begin{equation}
{\cal L}_{scalebreaking} =  -\frac{1}{4} \chi^{4} {\rm {ln}}
\Bigg ( \frac{\chi^{4}} {\chi_{0}^{4}} \Bigg ) + \frac{d}{3}{\chi ^4} 
{\rm {ln}} \Bigg ( \bigg ( \frac { \sigma^{2} \zeta }{\sigma_{0}^{2} 
\zeta_{0}}\bigg) \bigg (\frac {\chi}{\chi_0}\bigg)^3 \Bigg ).
\label{scalebreak}
\end{equation}
The Lagrangian density corresponding to the dilaton field, $\chi$
leads to the trace of the energy momentum tensor as 
\cite{heide1,chqsram}
\begin{equation}
\theta_{\mu}^{\mu} = \chi \frac{\partial {\cal L}}{\partial \chi} 
- 4{\cal L} 
= -(1-d)\chi^{4}.
\label{tensor1}
\end{equation}

The comparison of the trace of the energy momentum tensor arising
from the trace anomaly of QCD with that of the present chiral model
given by equation (\ref{tensor1}),
gives the relation of the dilaton field to the scalar gluon condensate.
We have, in the limit of finite quark masses \cite{cohen},
\begin{equation}
T_{\mu}^{\mu} = \sum_{i=u,d,s} m_i \bar {q_i} q_i+ \langle \frac{\beta_{QCD}}{2g} 
G_{\mu\nu}^{a} G^{\mu\nu a} \rangle  \equiv  -(1 - d)\chi^{4}, 
\label{tensor2m}
\end{equation}
where the first term of the energy-momentum tensor, within the chiral 
SU(3) model is the negative of the explicit chiral symmetry breaking
term, ${\cal L}_{SB}$. In the mean field approximation, this 
chiral symmetry breaking term is given as 
\begin{eqnarray}
{\cal L} _{SB} & = & {\rm Tr} \left [ {\rm diag} \left (
-\frac{1}{2} m_{\pi}^{2} f_{\pi} (\sigma+\delta), 
-\frac{1}{2} m_{\pi}^{2} f_{\pi} (\sigma-\delta), 
\Big( \sqrt{2} m_{k}^{2}f_{k} - \frac{1}{\sqrt{2}} 
m_{\pi}^{2} f_{\pi} \Big) \zeta \right) \right ]. 
\label{ecsb}
\end{eqnarray}

In the above, we have explicitly written down the matrix 
whose trace gives the Lagrangian density corresponding to  
the explicit chiral symmetry breaking in the chiral SU(3) model. 
Comparing the above term with the explicit chiral symmetry 
breaking term of the Lagrangian density in QCD given as
\begin{eqnarray}
{\cal L}^{QCD}_{SB} & =- & {\rm Tr} \left [ {\rm diag} \left (m_u \bar u u, 
m_d \bar d d , m_s \bar s s \right ) \right],
\label{ecsbqcd}
\end{eqnarray}
we obtain the nonstrange quark condensates ($\langle \bar u u \rangle$ and 
$\langle \bar d d \rangle$) and the strange quark condensate 
($\langle \bar s s \rangle $)  to be related to the
the scalar fields, $\sigma$, $\delta$ and $\zeta$
as 

\begin{equation}
m_u\langle \bar u u \rangle 
= \frac{1}{2}m_{\pi}^{2} f_{\pi} (\sigma+\delta)
\label{nsubu}
\end{equation}
\begin{equation}
m_d \langle \bar d d \rangle
= \frac{1}{2}m_{\pi}^{2} f_{\pi} (\sigma-\delta)
\label{nsdbd}
\end{equation}
and,
\begin{equation}
m_s\langle \bar s s \rangle 
= \Big( \sqrt {2} m_{k}^{2}f_{k} - \frac {1}{\sqrt {2}} 
m_{\pi}^{2} f_{\pi} \Big) \zeta.
\label{sbs}
\end{equation}

The coupled equations of motion for the non-strange scalar isoscalar 
field $\sigma$, scalar isovector field, $\delta$, the
strange scalar field $ \zeta$, and the dilaton 
field $\chi$, derived from the Lagrangian density,
are solved to obtain the values of these fields
in the strange hadronic medium. 

The QCD $\beta$ function occurring in the right hand side of equation
(\ref{tensor2m}), at one loop level, for 
$N_{c}$ colors and $N_{f}$ flavors, is given as
\begin{equation}
\beta_{\rm {QCD}} \left( g \right) = -\frac{11 N_{c} g^{3}}{48 \pi^{2}} 
\left( 1 - \frac{2 N_{f}}{11 N_{c}} \right)  +  O(g^{5})
\label{beta}
\end{equation}
We then obtain the trace of the energy-momentum tensor in QCD, 
using the one loop beta function given by equation (\ref{beta}),
for $N_c$=3 and $N_f$=3, as given by,
\begin{equation}
\theta_{\mu}^{\mu} = - \frac{9}{8} \frac{\alpha_{s}}{\pi} 
G_{\mu\nu}^{a} G^{\mu\nu a}
+ \left( m_{\pi}^{2} 
f_{\pi} \sigma
+ \Big( \sqrt {2} m_{k}^{2}f_{k} - \frac {1}{\sqrt {2}} 
m_{\pi}^{2} f_{\pi} \Big) \zeta \right), 
\label{tensor4m}
\end{equation} 
where $\alpha_s=\frac{g^2}{4\pi}$. Using equations (\ref{tensor2m}) 
and (\ref{tensor4m}), we can write  
\begin{equation}
\left\langle  \frac{\alpha_{s}}{\pi} {G^a}_{\mu\nu} {G^a}^{\mu\nu} 
\right\rangle =  \frac{8}{9} \Bigg [(1 - d) \chi^{4}
+ \left( m_{\pi}^{2} f_{\pi} \sigma
+ \Big( \sqrt {2} m_{k}^{2}f_{k} - \frac {1}{\sqrt {2}} 
m_{\pi}^{2} f_{\pi} \Big) \zeta \right) \Bigg ]. 
\label{chiglu}
\end{equation}
Hence the scalar gluon condensate of QCD ($\langle {G^a}_{\mu \nu}
G^{\mu \nu a} \rangle$) is simulated by a scalar dilaton field in the present
hadronic model. For the case of massless quarks, the scalar gluon condensate 
is proportional to the fourth power of the dilaton field, whereas for the
case of finite masses of quarks, there are modifications arising from 
the scalar fields, $\sigma$ and $\zeta$.

We calculate the light quark condensates, $\langle \bar u u\rangle$, 
$\langle \bar d d\rangle$ and $\langle \bar  s s \rangle$ 
and the scalar gluon condensate, 
$\left\langle  \frac{\alpha_{s}}{\pi} {G^a}_{\mu\nu} {G^a}^{\mu\nu} 
\right\rangle$ in the hadronic medium  using 
the equations (\ref{nsubu}), (\ref{nsdbd}) and (\ref{sbs}) and (\ref{chiglu}) 
respectively, from the medium modifications of the scalar fields, $\sigma$, 
$\delta$, $\zeta$ and $\chi$. These values of the quark and gluon condensates
are then taken as inputs for the studying the masses of the light vector mesons
($\omega$, $\rho$, $\phi$) in the strange hadronic matter 
using the QCD sum rule approach. In the next section we shall describe
the QCD sum rule approach to study these in-medium vector meson masses 
in the isospin asymmetric strange hadronic medium.

\section{QCD sum rule approach}
In the present section, we investigate the properties of the
light vector mesons ($\omega$, $\rho$, $\phi$) in the nuclear medium 
using the method of QCD sum rules. The in-medium masses of the vector mesons
are computed from the medium modifications of the light quark condensates
and the scalar gluon condensate calculated in the chiral effective
model as described in the previous section.
The current current correlation function for the
vector meson, V(=$\omega$,$\rho$, $\phi$) is written as

\begin{equation}
\Pi _{\mu \nu}= i\int d^4x d^4y \langle 0| T j^V_\mu (x) j^V _\nu (0)|0\rangle,
\end{equation}
where $T$ is the time ordered product and $J^V_\mu$ is the 
current for the vector meson, $V=\rho,\omega,\phi$, given as
$j_\mu ^{\rho}=\frac{1}{2} 
(\bar u \gamma_\mu u -\bar d \gamma_\mu d)$,
$j_\mu ^{\omega}=\frac{1}{6} 
(\bar u \gamma_\mu u +\bar d \gamma_\mu d)$ and
$j_\mu ^{\phi}=-\frac{1}{3} (\bar s \gamma_\mu s)$.
Current conservation gives the transverse tensor structure for the
correlation function as
\begin{equation}
\Pi^V_{\mu \nu}(q)=\left (g_{\mu \nu}-\frac{q_\mu q_\nu}{q^2}
\right) \Pi^V (q^2)
\end{equation}
where, 
\begin{equation}
\Pi^V (q^2)=\frac{1}{3} g^{\mu \nu}\Pi^V _{\mu \nu }(q).
\end{equation}
The correlation function $\Pi^V (q^2)$ in the large space-like region
$Q^2=-q^2 >> $ 1 GeV$^2$ for the light vector mesons ($\omega$, $\rho$
and $\phi$) can be written in terms of the operator product
expansion (OPE) as \cite{klinglnpa,kwonprc2008}

\begin{equation}
12\pi^2{\tilde \Pi^V} (q^2=-Q^2)=
d_V \Big [ -c^V_0 \ln \Big (\frac { Q^2}{\mu^2}\Big )
+\frac {c^V_1}{Q^2} + \frac {c^V_2}{Q^4} +\frac {c^V_3}{Q^6}+\cdots \Big ]
\label{qcdope}
\end{equation}
where, $\tilde \Pi^V (q^2=-Q^2)=\frac{ \Pi^V (q^2=-Q^2)}{Q^2}$ and
$\mu$ is a scale which we shall take as 1 GeV in the present
investigation. The coefficients $c^V_i$'s in equation (\ref{qcdope})
contain the informations of the nonperturbative effects
of QCD in terms of the quark and gluon condensates.
In equation (\ref{qcdope}), $d_V$=3/2,1/6 and 1/3,
for $\rho$, $\omega$ and $\phi$ vector mesons respectively.

For the vector mesons, $\rho$ and $\omega$, containing the
u and d quarks (antiquarks), these coefficients are
given as \cite{klinglnpa}
\begin{equation}
c_0 ^{(\rho,\omega)}=1+\frac{\alpha_s (Q^2)}{\pi},\;\;\;\;
c_1 ^{(\rho,\omega)}=-3 (m_u ^2 +m_d ^2)
\label{c0c1rhomg}
\end{equation}
\begin{equation}
c_2 ^{(\rho,\omega)}= \frac {\pi^2}{3}
\langle \frac {\alpha_s}{\pi} G^{\mu \nu} G_{\mu \nu}
\rangle + 4\pi^2 \langle m_u \bar u u +m_d \bar d d \rangle
\label{c2rhomg}
\end{equation}
\begin{eqnarray}
{c_3}^{(\rho,\omega)} & =& -4\pi^3 \Big [ \langle \alpha_s 
(\bar u \gamma_\mu \gamma_5 \lambda^a u 
\mp \bar d \gamma_\mu \gamma_5 \lambda^a d )^2 \rangle
\nonumber \\
&+& \frac {2}{9} \langle \alpha_s 
(\bar u \gamma_\mu \lambda^a u 
+ \bar d \gamma_\mu \lambda^a d )
(\sum_{q=u,d,s}\bar q \gamma^\mu \lambda^a q) \rangle \Big ]
\label {c3rhomg}
\end{eqnarray}
In the above, $\alpha_S =4\pi /(b \ln (Q^2/{\Lambda_{QCD}}^2))$
is the running coupling constant, with $\Lambda_{QCD}$=140MeV
and $b=11-(2/3)N_f$=9. 
In equation (\ref{c3rhomg}), the `$\mp$' sign in the first term
corresponds to $\rho (\omega)$ meson.

For $\phi$ meson, these coefficients are given as \cite{klinglnpa,svznpb1}
\begin{equation}
c_0 ^{\phi}=1+\frac{\alpha_s (Q^2)}{\pi},\;\;\;\;
c_1 ^{\phi}=-6 {m_s}^2 
\label{c0c1phi}
\end{equation}
\begin{equation}
c_2 ^{\phi}= \frac {\pi^2}{3}
\langle \frac {\alpha_s}{\pi} G^{\mu \nu} G_{\mu \nu}
\rangle + 8\pi^2 \langle m_s \bar s s \rangle
\label{c2phi}
\end{equation}
\begin{eqnarray}
{c_3}^{\phi}  = -8\pi^3 \Bigg [ 2 \langle \alpha_s 
(\bar s \gamma_\mu \gamma_5 \lambda^a s )^2 \rangle
+ \frac {4}{9} \langle \alpha_s 
(\bar s \gamma_\mu \lambda^a s )
(\sum_{q=u,d,s}\bar q \gamma^\mu \lambda^a q) \rangle \Bigg ]
\label{c3phi}
\end{eqnarray}
After Borel transformation, the correlator for the vector meson
given by equation (\ref{qcdope}) can be written as
\begin{equation}
 12 \pi^2 \tilde  \Pi^V (M^2)=d_V \Big [ c^V_0 M^2 +c^V_1+ \frac{c^V_2}{M^2}
+\frac{ c^V_3}{2M^4}\Big ]
\label{corropeborel}
\end{equation} 
On the phenomenological side, the correlator function,
$\tilde \Pi^V (Q^2)$ can be written as
\begin{equation}
12 \pi^2 \tilde \Pi^V _{phen}(Q^2)
=\int _0 ^\infty ds \frac{R^V_{phen}(s)}{s+Q^2}
\label{corrphen}
\end{equation}
where $R^V_{phen}(s)$ is the spectral density proportional to the
imaginary part of the correlator
\begin{equation}
R^V_{phen}(s)={12 \pi} {\rm {Im}} \Pi^V _{phen} (s).
\end{equation}
On Borel transformation, equation (\ref{corrphen}) reduces to 
\begin{equation}
12 \pi^2 \tilde \Pi^V (M^2)=\int _0 ^\infty d s e^{-s/{M^2} }
R^V_{phen}(s)
\label{corrphenborel}
\end{equation}
Equating the correlation functions from the phenomenological side
given by equation (\ref{corrphenborel}) to that from the operator
product expansion given by equation (\ref{corropeborel}),
we obtain,
\begin{equation}
\int _0 ^\infty d s e^{-{s}/{M^2} }
R^V_{phen} (s) ={d_V}
\Big [ c^V_0 M^2 +c^V_1+ \frac{c^V_2}{M^2}
+\frac{ c^V_3}{2M^4}\Big ].
\label{qsr}
\end{equation}
The finite energy sum rules (FESR) for the vector mesons are derived
from equation (\ref{qsr}) by assuming that the spectral density
separates to a resonance part ${R^V}_{phen}^{(res)}(s)$ with
$s \le s^V_0$ and a perturbative continuum as
\begin{equation}
R^V_{phen}(s) ={R^V}_{phen}^{(res)}(s) \theta (s^V_0-s)
+{d_V} c^V_0 \theta (s-s^V_0)
\label{qsr1}
\end{equation}
For $M > \sqrt {s^V_0}$, the exponential function in the integral of
the left hand side of the equation (\ref{qsr}) can be expanded 
in powers of $s/M^2$ for $s < s^V_0$. We then obtain the left hand side
of equation (\ref{qsr}) as
\begin{eqnarray}
&& \int _0 ^\infty e^{-s/M^2}R^V_{phen}(s)
=\int _0 ^{s^V_0} d s {R^V}_{phen}^{(res)} (s) -\frac {1}{M^2} 
\int _0 ^{s^V_0} d s s {R^V}_{phen}^{(res)} (s) + \frac{1}{2 M^4}
\int _0 ^{s^V_0} d s s^2 {R^V}_{phen} ^{(res)} (s) \nonumber \\
&+& 
{d_V} c_0  M^2 \Bigg (1- \frac{s^V_0}{M^2} +\frac{{(s^V_0)}^2}{2 M^4}
+\frac{{(s^V_0)}^3}{6 M^6}-\cdots \Bigg )
\label{rhophborel}
\end{eqnarray}
Equating the powers in $1/{M^2}$  in the Borel transformations of the
spectral functions, given by equations (\ref{qsr1}) and (\ref{rhophborel}),
we obtain the Finite energy sum rules (FESR) as
\begin{equation}
\int _0 ^{s^V_0} d s {R^V}_{phen}^{(res)} =
{d_V} (c^V_0 s^V_0 +c^V_1)  
\label{fesr1v}
\end{equation}

\begin{equation}
\int _0 ^{s^V_0} d s s {R^V}_{phen}^{(res)}=
{d_V} \Big (
\frac {(s^V_0)^2 c^V_0 }{2}-c^V_2 \Big )
\label{fesr2v}
\end{equation}

\begin{equation}
\int _0 ^{s^V_0} d s s^2 {R^V}_{phen}^{(res)}=
{d_V} \Big (
\frac{(s^V_0)^3}{3} c^V_0 +c^V_3 \Big )
\label{fesr3v}
\end{equation}
To evaluate $c^V_3$ for the vector mesons $\rho$, $\omega$ and $\phi$,
given by equations (\ref{c3rhomg}) and (\ref{c3phi}),
we use factorization method \cite{svznpb2}, 
\begin{equation}
\langle (\bar {q_i} \gamma_\mu \gamma_5 \lambda^a {q_j})^2 \rangle
= -\langle (\bar {q_i} \gamma_\mu \lambda^a {q_j})^2 \rangle
=\delta_{ij} \frac {16}{9} \kappa_i \langle \bar {q_i} {q_i} \rangle ^2,
\label{4qfact}
\end{equation}
for ${q_i}=u,d,s$ for $i=1,2,3$.
In the above, $\kappa_i$ is introduced 
to parametrise the deviation from exact factorization ($\kappa_i$=1).
Using equation (\ref{4qfact}), the four quark condensate for the
$\omega (\rho)$ meson given by equation (\ref{c3rhomg}) becomes
\begin{equation}
{c_3}^{(\rho,\omega)}=
-\alpha_s \pi^3\times \frac{448}{81} \kappa_q (\langle \bar u u \rangle^2
+ \langle \bar d d \rangle^2),
\label{c3rhomgf}
\end{equation}
where we have used, $\kappa_u \simeq \kappa_d =\kappa_q$. 

For the $\phi$ meson, using equations (\ref{c3phi}) and (\ref{4qfact}),
we obtain the four quark condensate, ${c_3}^\phi$ as
given by \cite{svznpb1}
\begin{eqnarray}
{c_3}^{\phi} 
&=& -8\pi^3 \times \frac{224}{81} \alpha_s \kappa_s 
\langle \bar s s \rangle ^2.
\label{c3phif}
\end{eqnarray}

We assume a simple ansatz for the spectral function, $R^V_{phen}(s)$
as \cite{klinglnpa,kwonprc2008}
\begin{equation}
R^V_{phen}(s)=F_V \delta (s-{m_V}^2)+ d_V c^V_0 \theta (s-s^V_0),
\label{spspectf}
\end{equation}

Using the form of the spectral function given by equation (\ref{spspectf}),
the finite energy sum rules  for vacuum given by equations (\ref{fesr1v}) 
to (\ref{fesr3v}), can be written as
\begin{equation}
F_V =
{d_V} (c^V_0 s^V_0 +c^V_1)  
\label{fesr1vf}
\end{equation}
\begin{equation}
F_V m_V^2=
{d_V} \Big (
\frac {(s^V_0)^2 c^V_0 }{2}-c^V_2 \Big )
\label{fesr2vf}
\end{equation}
\begin{equation}
F_V m_V^4=
{d_V} \Big (
\frac{(s^V_0)^3}{3} c^V_0 +c^V_3 \Big )
\label{fesr3vf}
\end{equation}
Using equations (\ref{fesr1vf}) and (\ref{fesr2vf}), we determine the values 
of $F_V$ and $s^V_0$ by assuming the values of $c^V_0$, with $Q^2=s_0$
($\alpha_s(Q^2 \simeq 1 {\rm GeV}^2)$=0.5 )
and $c^V_1$ as calculated in the chiral SU(3) model.
These values are assumed in equation (\ref{fesr3vf}) to find the 
vacuum value of the 4-quark condensate, $c^V_3$ and hence
the value of $\kappa_i$.

At finite densities, there is contribution to the spectral function
for the vector mesons, due to scattering from the baryons and 
the equation (\ref{qsr}) is modified to
\begin{equation}
\int _0 ^\infty d s e^{-{s}/{M^2} }
R^ V_{phen} (s)+12 \pi^2 \Pi^ V(0) ={d_V}
\Big [ c_0 M^2 +c_1+ \frac{c_2}{M^2}
+\frac{ c_3}{2M^4}\Big ],
\label{qsrfinitedens}
\end{equation}
where, in the nuclear medium, $\Pi^ V(0)=\frac {\rho_B}{4M_N}$ 
for V=$\omega$,$\rho$.  and vanishes for $\phi$ meson
\cite{klinglnpa,hatlee2,bochkarev,florkowski}. 
However, in the presence of hyperons in the hadronic medium, 
the contribution due to the scattering of the $\omega$ and 
$\rho$ vector mesons from the baryons is modified to 
\begin{equation}
\Pi^ V(0)=\frac{1}{4}\sum_i \Big (\frac{g_{Vi}}{g_{VN}}\Big )^2
\frac{\rho_i}{M_i},
\end{equation}
where, $g_{Vi}$ is the coupling strength of the
vector meson, V with the $i$-th baryon ($i=N,\Lambda,\Sigma^{\pm,0},
\Xi^{-,0}$), $\rho_i$ and $M_i$
are the number density and  mass of the $i$-th baryon. 
For the $\omega$ meson, 
$\frac{g_{\omega i}}{g_{\omega N}}
=(1,\frac{2}{3},\frac{2}{3},
\frac{1}{3})$ for $i=N,\Lambda,\Sigma^{\pm,0},\Xi^{-,0}$
respectively. For the $\rho$ meson, the ratio  
$\frac{g_{\rho i}}{g_{\rho N}} =(1,0,2,1)$
for $i=(N,\Lambda,\Sigma^{\pm,0},\Xi^{-,0})$.
In the nuclear medium, the contribution for the $\phi$ meson due to scattering 
from nucleons vanishes, since the $\phi$ meson-nucleon coupling strength
is zero. In the strange hadronic matter, the contribution is, however,
nonzero due to the presence of the hyperons in the medium.
For the $\phi$ meson, $\frac{g_{\phi i}}{g_{\phi \Lambda}} =(1,1,2)$
for $i=(\Lambda,\Sigma^{\pm,0},\Xi^{-,0})$.

At finite densities, the finite energy sum rules (FESR) for vacuum 
given by equations (\ref{fesr1vf}) to (\ref{fesr3vf}) are modified to
\begin{equation}
F^*_V =
{d_V} ({c^V_0} {{s^*}^V_0} +{c^V_1}) -12\pi^2 \Pi^V(0) 
\label{fesr1mf}
\end{equation}
\begin{equation}
F^*_V {m^*_V}^2=
{d_V} \Big (
\frac {({s^*}^V_0)^2 c^V_0}{2}-{c^*}_2^V \Big )
\label{fesr2mf}
\end{equation}
\begin{equation}
F^*_V {m^*_V}^4=
{d_V} \Big (
\frac{({s^*}^V_0)^3}{3} c^V_0 +{c^*}_3^V \Big )
\label{fesr3mf}
\end{equation}
These equations are solved to obtain the medium dependent mass,
$m^*_V$, the scale ${s^*}_0^V$ and $F^*_V$, by using the coefficient
$k$ of the 4-quark condensate for the vector mesons,
as determined from the  FESRs in vacuum.
\begin{figure}
\includegraphics[width=16cm,height=16cm]{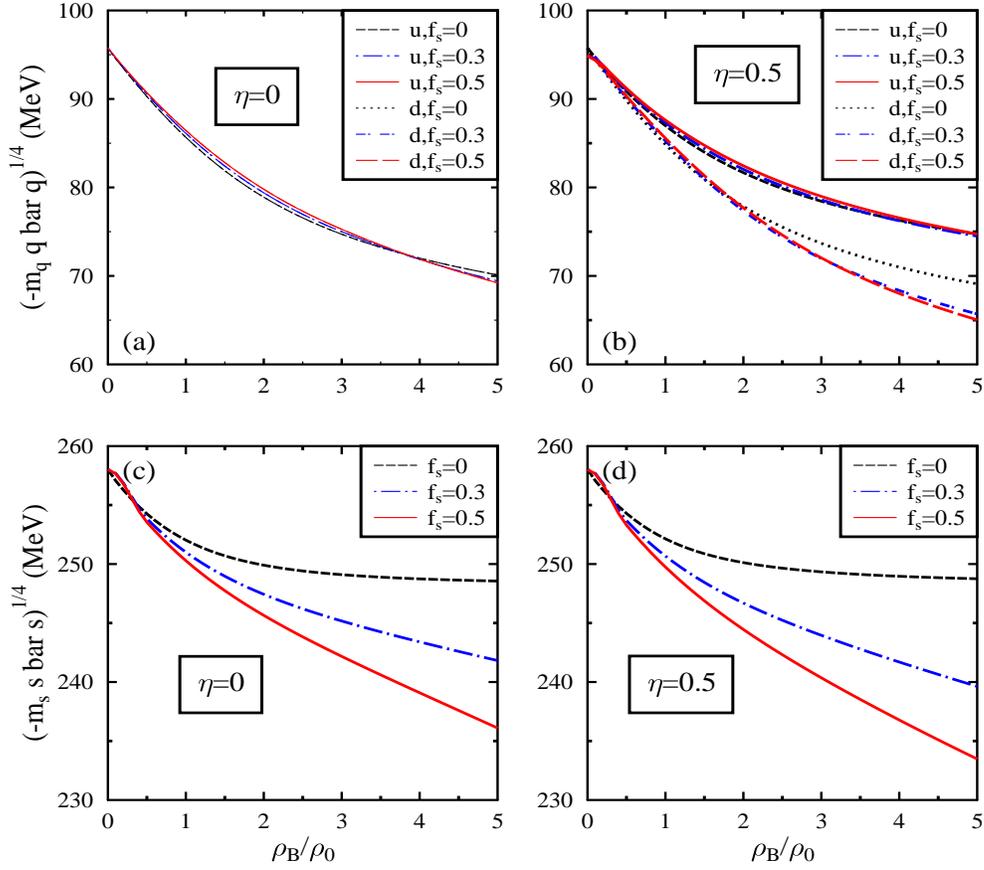}
\caption{(Color online)
The quark condensates $(-m_q \langle \bar q q \rangle)^{1/4}$
($q=u,d$) and $(-m_s \langle \bar s s \rangle)^{1/4}$, 
in units of MeV, are plotted as
functions of density for isospin asymmetric hadronic matter
(for $f_s$=0, 0.3 and 0.5) in figures (b) and (d),
and compared with the isospin symmetric case, 
shown in subplots (a) and (c).
}
\label{psipsibdens}
\end{figure}
\begin{figure}
\includegraphics[width=16cm,height=16cm]{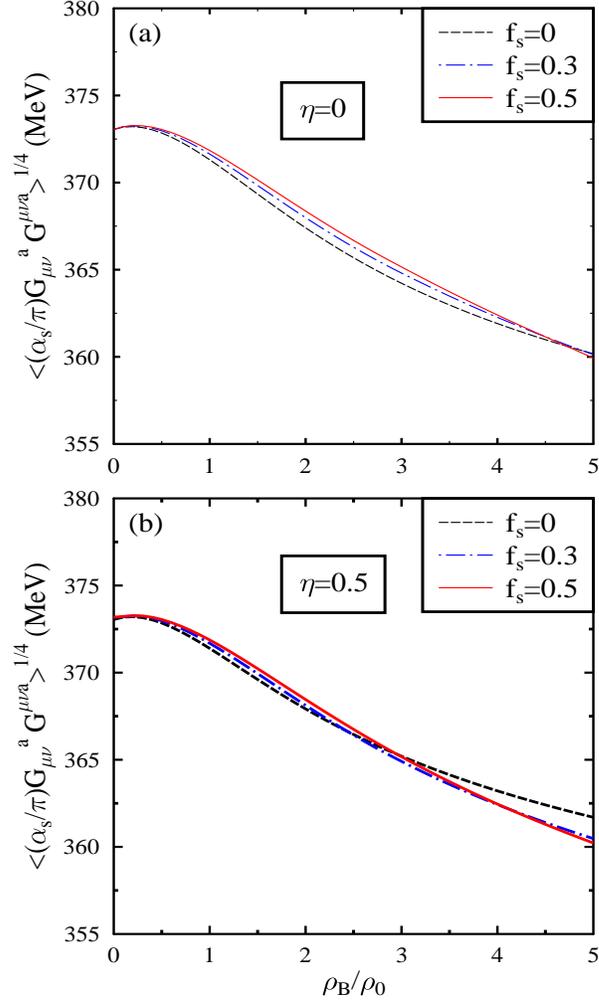}
\caption{(Color online)
The quantity
$\langle  \frac{\alpha_{s}}{\pi} {G^a}_{\mu\nu} {G^a}^{\mu\nu} 
\rangle^{1/4}$ in MeV plotted as a function of the baryon density in units
of the nuclear matter saturation density. This is plotted for isospin 
asymmetric hadronic matter (for strangeness fraction, $f_s$=0, 0.3, 0.5 and
isospin asymmetric parameter, $\eta$=0.5) in subplot (b) and compared with
the symmetric matter ($\eta$=0) in subplot (a).} 
\label{ggcond}
\end{figure}
\begin{figure}
\includegraphics[width=16cm,height=16cm]{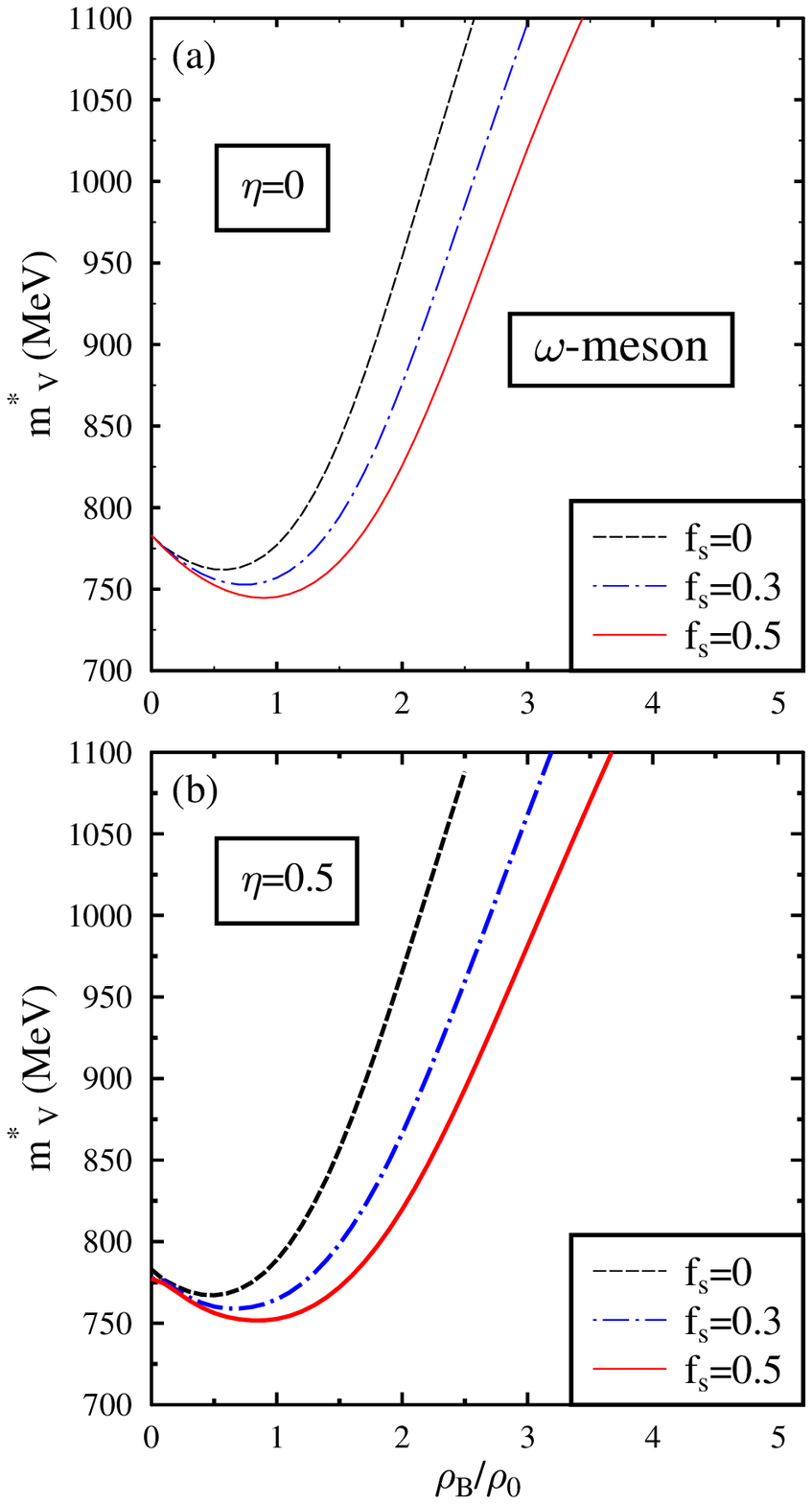}
\caption{(Color online)
The mass of $\omega$ meson plotted as a function of the
baryon density in units of nuclear saturation density,
for the isospin asymmetric strange hadronic matter
(for strangeness fraction, $f_s$=0, 0.3, 0.5 and
isospin asymmetric parameter, $\eta$=0.5) in subplot (b) and compared with
the symmetric matter ($\eta$=0) shown in subplot (a).} 
\label{omgmassdens}
\end{figure}
\begin{figure}
\includegraphics[width=16cm,height=16cm]{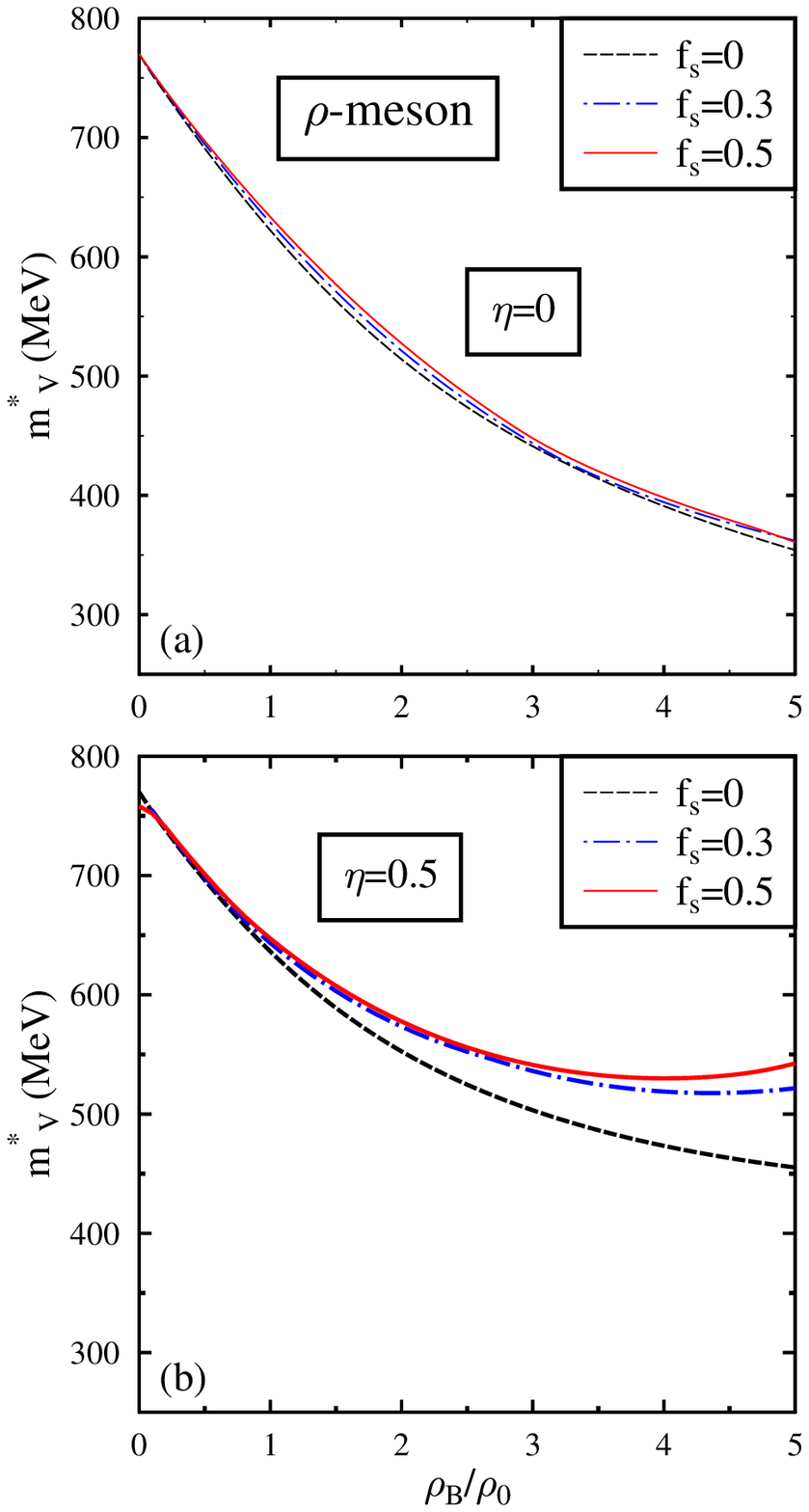}
\caption{(Color online)
The mass of $\rho$ meson plotted as a function of the
baryon density in units of nuclear saturation density,
for the isospin asymmetric strange hadronic matter
(for strangeness fraction, $f_s$=0, 0.3, 0.5 and
isospin asymmetric parameter, $\eta$=0.5) in subplot (b) and compared with
the symmetric matter ($\eta$=0) shown in subplot (a).} 
\label{rhomassdens}
\end{figure}
\begin{figure}
\includegraphics[width=16cm,height=16cm]{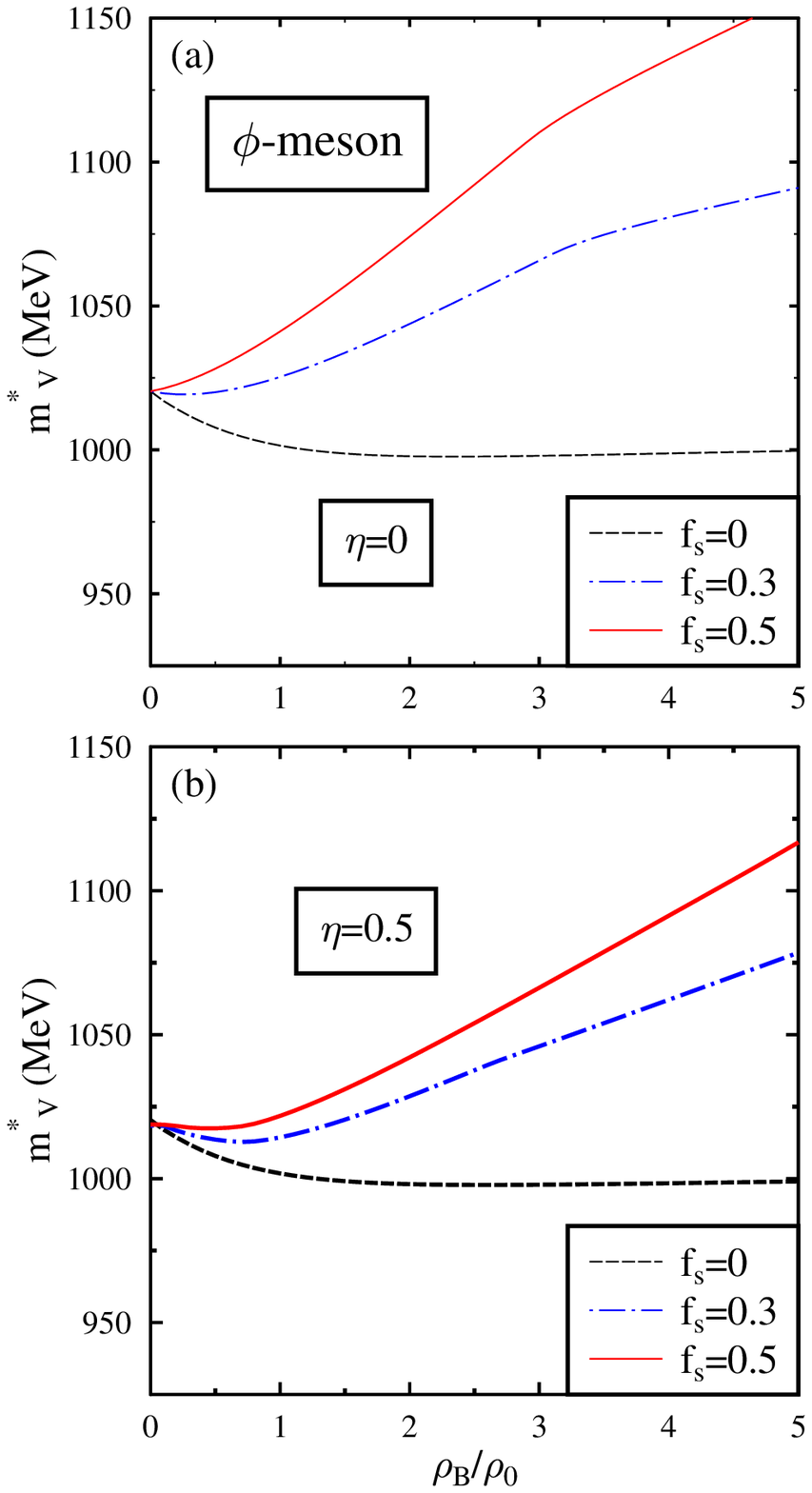}
\caption{(Color online)
The mass of $\phi$ meson plotted as a function of the
baryon density in units of nuclear saturation density,
for the isospin asymmetric strange hadronic matter
(for strangeness fraction, $f_s$=0, 0.3, 0.5 and
isospin asymmetric parameter, $\eta$=0.5) in subplot (b) 
and compared with the symmetric matter ($\eta$=0) 
shown in subplot (a).} 
\label{phimassdens}
\end{figure}
\begin{figure}
\includegraphics[width=16cm,height=16cm]{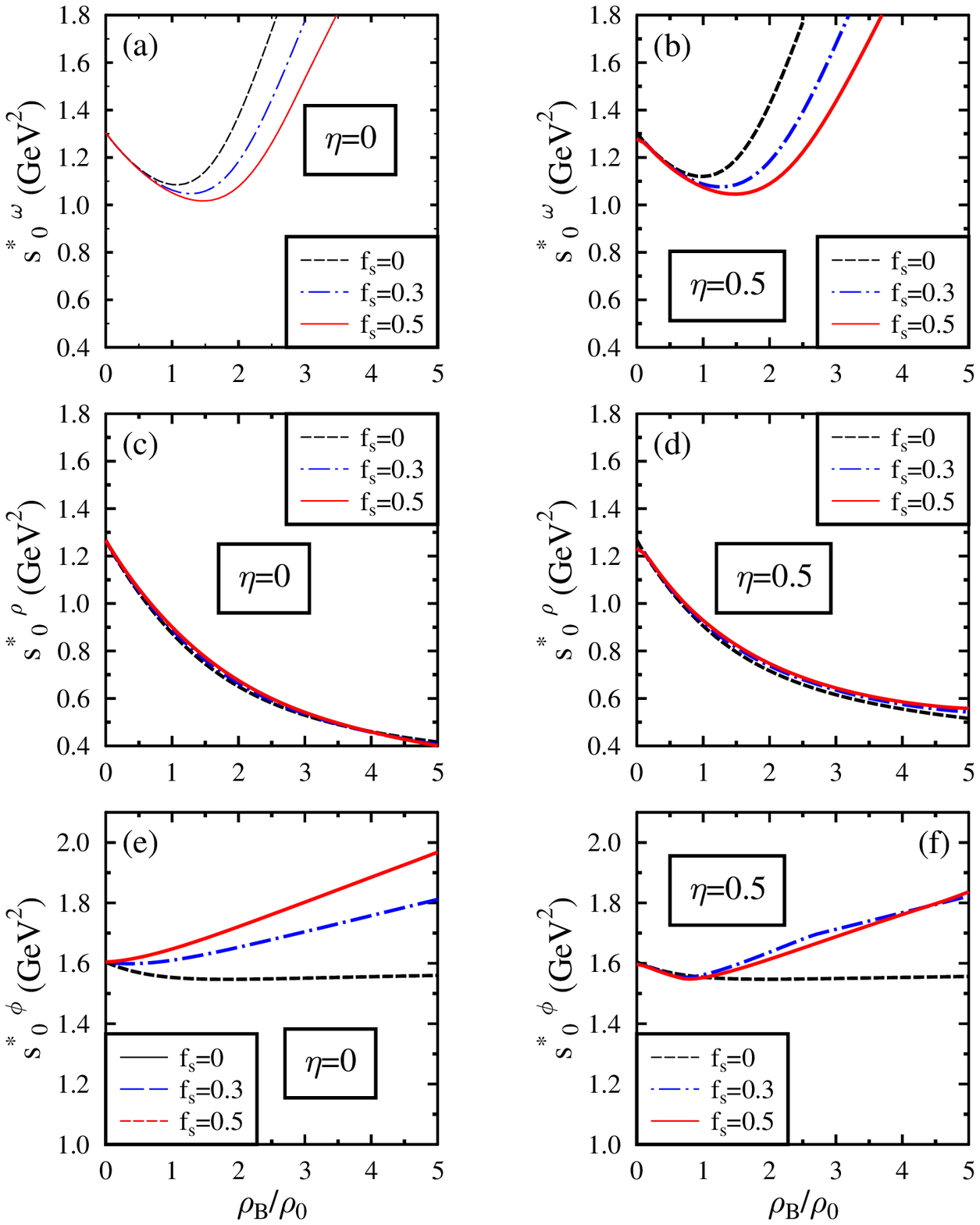}
\caption{(Color online)
The density dependence of ${s^*}_0^V$ for the vector mesons 
($\omega$, $\rho$ and $\phi$) in the strange hadronic matter 
is shown for the isospin symmetric ($\eta$=0) and isospin asymmetric 
(with $\eta$=0.5) cases for values of $f_s$=0, 0.3 and 0.5.
}
\label{s0dens}
\end{figure}
\begin{figure}
\includegraphics[width=16cm,height=16cm]{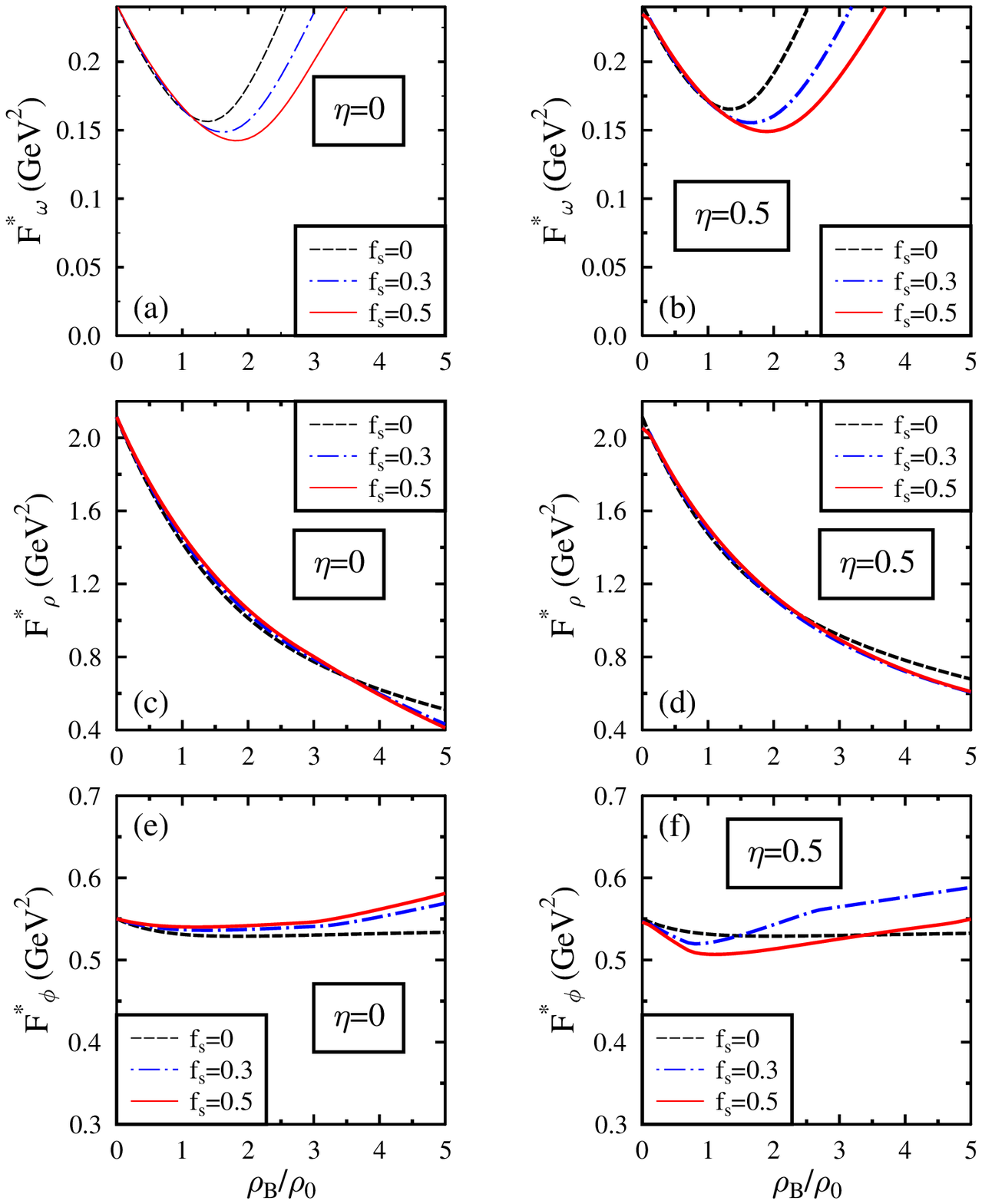}
\caption{(Color online)
The density dependence of $F^*_V$ for the vector mesons 
($\omega$, $\rho$ and $\phi$) in the strange
hadronic matter is shown for the isospin symmetric ($\eta$=0) 
and isospin asymmetric 
(with $\eta$=0.5) cases for values of $f_s$=0, 0.3 and 0.5.
}
\label{fvdens}
\end{figure}
\begin{figure}
\includegraphics[width=16cm,height=16cm]{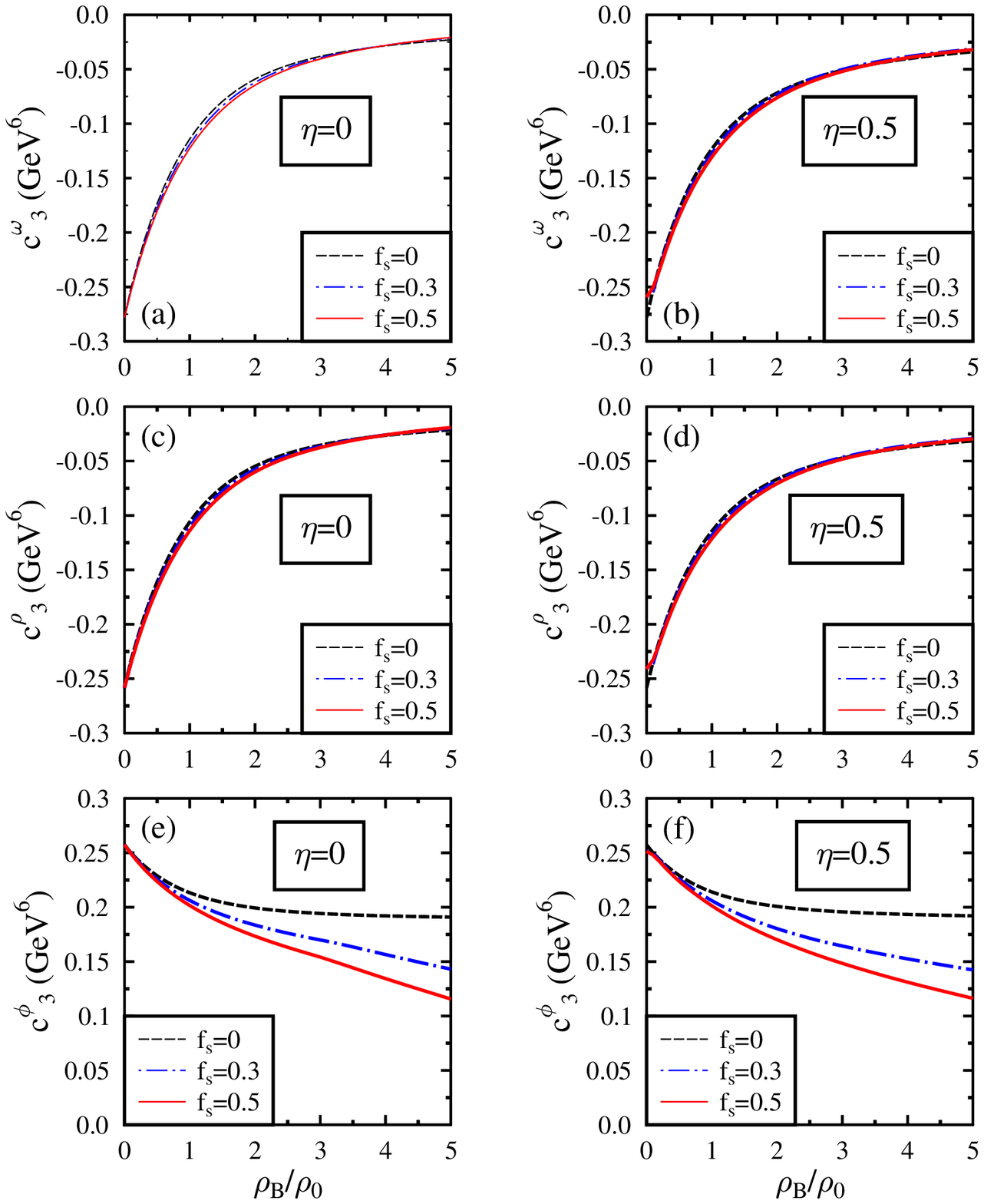}
\caption{(Color online)
The density dependence of the 4-quark condensate for the cases of
the $\omega$, $\rho$ and $\phi$ mesons is shown for the isospin 
symmetric ($\eta$=0) and asymmetric (with $\eta$=0.5) hadronic matter 
for the values of the strangeness fraction, $f_s$=0, 0.3 and 0.5.
}
\label{4quark}
\end{figure}
\section{Results and Discussions}

In this section, we first investigate the effects of density 
on the scalar gluon condensate and the light quark condensates
arising due to the modifications of the dilaton field, $\chi$
and the scalar isoscalar fields $\sigma$ and $\zeta$ calculated in
the chiral SU(3) model. The values of these fields in the isospin
asymmetric strange hadronic matter are obtained by solving the
coupled equations of these fields in the mean field approximation.
The nonstrange and strange quark condensates, 
$\langle \bar q q\rangle $ ($q=u,d$)
and $\langle \bar s s \rangle $, as well as,
the scalar gluon condensate, 
$\left\langle  \frac{\alpha_{s}}{\pi} {G^a}_{\mu\nu} {G^a}^{\mu\nu} 
\right\rangle$,
are calculated from the in-medium values of the fields $\sigma$, $\zeta$
and $\chi$, by using the equations  (\ref{nsubu}), (\ref{nsdbd}),
(\ref{sbs}) and (\ref{chiglu}), respectively. The values of the current
quark masses are taken as  $m_u$=4MeV, $m_d$=7MeV and $m_s$=150MeV 
in the present 
investigation. In figure \ref{psipsibdens}, we show the density dependence
of the quantities $(-m_q\langle \bar q q\rangle)^{1/4} $ ($q=u,d$), 
$(-m_s \langle \bar s s\rangle)^{1/4} $ for given isospin asymmetry and
strangeness of the hadronic medium. For the isospin symmetric situation
($\eta$=0), the quantity $(-m_q\langle \bar q q\rangle)^{1/4} $
is identical for $u$ and $d$ quarks, for a given value of $f_s$. 
It is also seen that the effect from the strangeness fraction is
very small. For the isospin asymmetric situation, the quantities
$(-m_u\langle \bar u u\rangle)^{1/4} $ and 
$(-m_d\langle \bar d d\rangle)^{1/4} $ are no longer identical,
and their difference is due to the nonzero value of the isoscalar scalar
field $\delta$, as can be seen from equations (\ref{nsubu}) and 
(\ref{nsdbd}). For the $u$ quark, there is seen to be smaller drop 
with density as compared to the $d$ quark due to the negative value 
of the isoscalar scalar field $\delta$ in the medium. 
For the isospin symmetric nuclear matter, the value of the quantity
$(-m_q\langle \bar q q\rangle)^{1/4} $ for $q=u,d$, changes from
the vacuum value of 95.8 MeV to 85.7 MeV at the nuclear matter saturation
density. This corresponds to a drop of the quantity, 
$(-m_q\langle \bar q q\rangle) $ for $q=u,d$,
by about 36 \%  at the nuclear matter saturation
density from its vacuum value. At a density of 4$\rho_0$,
this quantity is modified to (72 MeV)$^4$, which corresponds to 
a drop of about 69 \% from its vacuum value. The drop of 
the non-strange condensate 
in the medium is the dominant contribution to the modification 
of the $\omega$ and $\rho$ mesons in the medium.
The quantity $(-m_s \langle \bar s s\rangle)^{1/4} $ 
for given isospin symmetric ($\eta$=0) and isospin asymmetric 
(with $\eta$=0.5) situations are shown in subplots 
(c) and (d) respectively. The vacuum value of 
$(-m_s \langle \bar s s\rangle)^{1/4} $ is about 258 MeV, 
which may be compared with the value of 210 MeV 
in Ref. \cite{hatlee}. 
For the symmetric nuclear matter, the quantity 
$(-m_s \langle \bar s s\rangle)^{1/4} $ changes from the vacuum
value of 258 MeV to 252 MeV and 248.7 MeV at densities of
$\rho_0$ and 4$\rho_0$ respectively, which correspond to 
about 9\%  and 13.7 \% drop in the
quantity, 8$\pi^2 \langle m_s \bar s s \rangle$ occurring 
in $c_2 ^\phi$ in the finite energy sum
rule for the $\phi$ meson given by equation (\ref{c2phi}).
Figure \ref{ggcond} shows the 
quartic root of the scalar gluon condensate, 
$\langle  \frac{\alpha_{s}}{\pi} {G^a}_{\mu\nu} {G^a}^{\mu\nu} 
\rangle^{1/4}$ as a function of the baryon density in units
of the nuclear matter saturation density, for isospin symmetric
($\eta$=0) as well as
asymmetric hadronic medium (with $\eta$=0.5) for typical values 
of the strangeness fraction. The value of the scalar gluon condensate 
$\langle \frac{\alpha_{s}}{\pi} {G^a}_{\mu\nu} {G^a}^{\mu\nu} 
\rangle$ for isospin symmetric nuclear matter is observed to be
modified from the vacuum value of (373 MeV)$^4$ to (371.3 MeV)$^4$ 
and (361.9 MeV)$^4$ at densities of $\rho_0$ and 4$\rho_0$ respectively, 
which correspond to about 1.8\% and 11.4\% drop in the medium
from its vacuum value. It is thus
observed that the light nonstrange quark condensate has a larger drop in the
medium as compared to the strange quark condensate as well as
the scalar gluon condensate in the medium. This is observed as 
a much smaller drop of the $\phi$ meson in the medium
as compared to the $\omega$ and $\rho$ mesons due to the
quark and gluon condensates. The in-medium quark 
and gluon condensates are used as inputs 
for the calculations of the vector meson masses in the hadronic
medium. The effects of isospin asymmetry as well as strangeness
of the medium on the masses of the vector mesons
are investigated in the present work. 
As has already been mentioned, using the vacuum values of the
vector meson mass and the quark and gluon condensates, 
the finite energy sum rules (FESR) for the vector mesons in vacuum
given by equations (\ref{fesr1vf}), (\ref{fesr2vf}) and
(\ref{fesr3vf}) are solved to obtain the values for $s_0^V$, $F_V$ 
and the coefficient of the 4-quark condensate, $\kappa_{q(s)}$.  
The vacuum value of the scale, $s_0^V$, which separates the resonance 
part from the continuum part is obtained as 1.3 ,1.27 and 1.6 GeV$^2$
and the value of $F_V$ is obtained as  0.242,0.258 and 0.55 GeV$^2$
for the for $\omega$, $\rho$ and $\phi$ mesons respectively. 
The value of the coefficient of the 4-quark condensate
is obtained as 7.788, 7.236 and -1.21 for the  $\omega$, $\rho$
and $\phi$ mesons, which are then used to obtain the medium
dependent mass, $m^*_V$, the scale ${s^*}_0^V$ and $F^*_V$
for the vector mesons,
by solving the FESRs in the medium given by equations
(\ref{fesr1mf}), (\ref{fesr2mf}) and (\ref{fesr3mf}).

In figure \ref{omgmassdens},
the density dependence of the mass of the $\omega$-meson is 
shown for the cases of isospin symmetric ($\eta$=0) as well as 
the asymmetric matter for given values of the strangeness fraction, 
$f_s$. There is seen to be initially a drop in the $\omega$-meson 
mass with increase in density. However, as the density is further 
increased, the mass of the $\omega$-meson is observed to increase 
with density. This behavior can be understood from the 
equations (\ref{fesr1mf}) and (\ref{fesr2mf}), which yield the
expression for the mass squared of the vector meson as
\begin{equation}
{m^*_V}^2= \frac{ \Big (
\frac {({s^*}^V_0)^2 c^V_0}{2}-{c^*}_2^V \Big )}
{({c^V_0} {{s^*}^V_0} +{c^V_1}) -{12\pi^2 (\Pi^V(0)}/{d_V})} 
\label{mv2}
\end{equation}
The contribution of $c_1^V$ is negligible for the $\rho$ and $\omega$ 
mesons, due to the small values of the masses of the $u$ and $d$ quarks.
At low densities, the contribution from the scattering of the vector 
mesons from baryons, given by the last term in the denominator
of (\ref{mv2}) is negligible and the mass drop of the $\omega$ meson
mainly arises due to the drop of the light quark condensates 
in the medium, given by the second term, ${c^*}_2^V$ in the numerator
which comes with a negative sign.
As seen in figure \ref{ggcond}, the modification of the
scalar gluon condensate of the term ${c^*}_2^V$ is much smaller
than that of the light quark condensate. However, at higher baryon 
densities, the last term in the denominator, the so-called Landau 
scattering term, becomes important for the $\omega$ meson. This leads to
an increase in the mass of the $\omega$ vector meson with density, as can be 
observed in figure \ref{omgmassdens}. The denominator becomes negative
above a certain value of density, when there does not exist any
solution for the mass of the $\omega$ meson, since ${m^*}_V^2$
becomes negative. For the case of nuclear matter, the mass
of the $\omega$ meson remains very similar in the isospin symmetric
as well as isospin asymmetric cases. This is because the modification
of the $\omega$ meson at low densities is mainly due to the quark 
condensates in the combination $(m_u \bar u u +m_d \bar d d$),
which depends only on the value of $\sigma$ (as seen from equations
(\ref{nsubu}) and (\ref{nsdbd})), and, $\sigma$ 
is marginally different for the symmetric and asymmetric cases. 
At higher densities, the effect of the Landau scattering term
becomes important. However, there is still observed to be very 
small difference
between the $\eta=0$ and $\eta$=0.5 cases of nuclear matter, 
since the dependence
of this term on the proton and neutron densities is in the form
($\rho_p$+$\rho_n$), which is same for the two cases at a given
density. 
With the inclusion of hyperons in the medium, the contribution of 
the scattering term in the denominator of equation (\ref{mv2}),
becomes smaller in magnitude due to 
the smaller values of the baryon-$\omega$ meson coupling strengths
for the hyperons as compared to the nucleons. However, the trend
of the initial mass drop followed by an increase at higher densities
is still seen to be the case for the mass of the $\omega$-meson. 
However, the
density above which the $\omega$-mass is observed to increase with 
density, is seen to be higher for finite strangeness fraction in the 
hadronic medium, since the contribution from the Landau scattering
term is smaller for the case of hyperonic matter as compared
to nuclear matter. For the $\rho$ meson,
the contribution from the Landau scattering term remains small
as compared to the contribution from the light quark condensate
in the medium, due to the factor $(1/{d_V})$ in this term,
which makes the contribution of the Landau scattering term 
to be 9 times smaller than that of the $\omega$ meson,
as ${(1/d_\rho)}/{(1/d_\omega)}=9$. 
This is observed as a monotonic decrease of the mass
of the $\rho$-meson with density, in figure \ref{rhomassdens}.
The effects of the strangeness fraction as well as isospin asymmetry
of the medium are seen to be small on the $\rho$ meson mass.
In figure \ref{phimassdens}, the mass of $\phi$ meson
is plotted as a function of the baryon density in units of nuclear
matter saturation density, for isospin symmetric and asymmetric cases
for typical values of the strangeness fraction. Due to the larger
value of the strange quark mass as compared to the $u(d)$ quark masses,
the contribution from $c_1^\phi$ is no longer negligible as
was the case for $\omega (\rho)$ meson. The dominant contribution
to the mass modification of the $\phi$ meson is from the in-medium
modification of the strange quark condensate of the Wilson coefficient
$c_2^\phi$ in the nuclear medium. This is because the $\phi$-meson 
has no contribution from the scattering term in nuclear matter,
since the nucleon-$\phi$ meson coupling is zero. The strange quark 
condensate  as well as the scalar gluon condensate have very small
effects from isospin asymmetry, leading to the modifications
of the $\phi$ meson mass to be very similar in the isospin symmetric
and asymmetric nuclear matter. There are, however, 
contributions from the Landau scattering term due to the hyperons 
in the medium for nonzero $f_s$, which leads to an increase
in the mass of the $\phi$ meson at higher values of the densities. 
For nuclear matter, the mass of the $\phi$ meson does not
have contribution from the scattering term and since the in-medium
modifications of both the strange quark condensate as well as 
the scalar gluon condensate are small and occur with opposite signs
in the coefficient ${c^*}_2^\phi$, the mass of $\phi$ meson
is observed to have negligible change with density, the value
being modified from the vacuum value of 1020 MeV to about 1000 MeV 
at a density of 5$\rho_0$. For the case of isospin symmetric hyperonic 
matter, there is seen to be an increase in the mass of the 
$\phi$ meson at low densities, due to scattering from the 
$\Xi^-$ and $\Xi^0$, whose number densities are equal for this
$\eta$=0 case. The $\Sigma^+$, $\Sigma^-$ and $\Sigma^0$ 
(with equal number densities) start appearing at around 3$\rho_0$, 
when the number densities of the $\Xi^-$ and $\Xi^0$ show a downward 
trend with density. It is the overall contributions from the hyperons 
to the scattering term which leads to the observed increase 
in the mass of the $\phi$ meson in the strange hadronic medium
with $f_s$=0.3 and 0.5, shown in figure \ref{phimassdens}. 
For the isospin asymmetric hyperonic matter,
there is contribution from the $\Sigma^+$ and $\Xi^0$ for $\eta$=0.5
situation (but not from $\Sigma^{0,-}$ and $\Xi^-$), which is seen 
as a smaller increase of the $\phi$ mass at high densities 
as compared to the isospin symmetric hyperonic matter. 

The density dependence of the scale ${s^*}_0^V$, which separates
the resonance part from the perturbative continuum, is
shown in figure \ref{s0dens} for the $\omega$, $\rho$ and $\phi$ 
vector mesons. For isospin symmetric nuclear matter, 
for the $\omega$ meson, the vacuum value of 1.3 GeV$^2$ 
is modified to about 1.086 and 1.375 GeV$^2$ at densities 
of $\rho_0$ and 2$\rho_0$ respectively. The dependence of ${s^*}_0^V$
on density as an initial drop followed by an increase
is similar to that of the density dependence of the  mass 
of the $\omega$ meson.
This can be understood in the following way. From the medium
dependent FESRs, we obtain the expression for the scale
${s^*}_0^V$ as 
\begin{equation}
{s^*}_0^V={m^*_V}^2\Bigg( 1+ \frac{2}{{m^*_V}^4c_0}
(c_1^V{{m^*}_V}^2+{c^*}_2^V-(12\pi^2 \Pi(0)/d_V))\Bigg)^{1/2}.
\end{equation} 
The value of the second term in the bracket, within the square root,
is found to be small as compared to 1. At higher densities,
the second term still remains small as compared to 1,
due to the cancelling effect of the contributions from the quark 
condensate and the Landau scattering term. 
This is seen as the density dependence of 
${s^*}_0^\omega$ to have first a drop and then an increase 
with density as found for the mass of the $\omega$ meson. 
The dependence of the scale ${s^*}_0^V$ for the $\rho$ meson
is observed to be a monotonic drop with increase in density, 
due to the negligible contribution from the Landau damping term
as compared to the contribution from the light quark condensate. 
In the case of $\phi$ meson, the effect of the scattering
term is zero for the nuclear matter case, when ${s^*}_0^\phi$
is observed to have a small drop due to the marginal drop 
of the strange condensate and the gluon condensate in the medium.
For the hyperonic matter, there is observed to be an increase
in ${s^*}_0^\phi$ due to the scattering from the hyperons,
which is observed to be larger for the isospin symmetric case
as compared to the isospin asymmetric situation.
In figure \ref{fvdens}, the value of $F^*_V$ is plotted
as a function of density. From the first finite energy sum rule
given by equation (\ref{fesr1mf}),
due to the small masses of the $u$ and $d$ quarks, the
term $c_1^V$ is negligible for the $\omega$ and $\rho$ mesons.
At low densities, the value of $F^*_V$ turns out to 
be proportional to ${s^*}_0^V$, since the contribution from
the Landau scattering term is small. At higher densities, there
is contribution from the Landau scattering term, which 
modifies the behavior of $F^*_V$ to a slower change with density
for the $\omega$ meson. For the $\rho$ meson, this is approximately
proportional to ${s^*}_0^\rho$ as the Landau term has negligible
contribution. For the $\phi$ meson, the scattering from the hyperons
leads to an increase of $F^*_\phi$ at higher densities.
In figure \ref{4quark}, we have plotted the quartic quark condensate,
$c_3^V$ for the $\omega$, $\rho$ and $\phi$ mesons, given
by equations (\ref{c3rhomgf}) and (\ref{c3phif})
as functions of density, for the isospin symmetric and asymmetric
nuclear (hyperonic) matter. For the $\rho$ and $\omega$ mesons,
the values of $\kappa$ calculated from the vacuum FESRs are 
found to be  7.236 and 7.788, which yield very similar values
for the 4-quark condensate for the $\omega$ and $\rho$ mesons,
shown in figure \ref{4quark}. The vacuum FESRs for the $\phi$ meson
yield the 4-quark condensate to be negative, with the value of
$\kappa$ as $-1.21$. There is seen to be a large effect from the strangeness
fraction of the medium on $c_3^\phi$, since the strange condensate
has appreciable effect from $f_s$, as can be seen from figure
\ref{psipsibdens}.

\section{Summary}
In the present investigation, we have calculated the effect
of density on the masses of the light vector mesons ($\omega$,
$\rho$ and $\phi$) using the QCD sum rules, from the light
quark condensates and gluon condensates in the medium calculated 
within a chiral SU(3) model. The effects of the isospin asymmetry 
as well as the strangeness of the medium on the modifications
of these masses have also been investigated. The light quark 
condensates ($\langle \bar u u\rangle $, 
$\langle \bar d d\rangle $, $\langle \bar s s\rangle $)
in the isospin asymmetric strange hadronic medium are calculated 
from the values of the nonstrange and strange scalar mesons,
$\sigma$ and $\zeta$, and the isoscalar scalar meson, $\delta$ of
the explicit symmetry breaking term of the chiral SU(3) model. 
The scalar gluon condensate is calculated from a scalar dilaton
field, which is introduced in the chiral SU(3) model to mimic 
the scale symmetry breaking of QCD. The mass of the $\omega$ meson 
is observed initially to drop with increase in density in the hadronic
matter. This is because the magnitudes of the light nonstrange quark 
condensates become smaller in the hadronic medium as compared 
to the values in vacuum. However, as the density 
is further increased, there is seen to be a rise in its  mass, 
when the effect from the Landau term due to the scattering of 
the $\omega$ meson from the baryons becomes important. 
In the presence of hyperons, the increase in the mass of the
$\omega$ meson occurs at a higher value of the density
as compared to the case of nuclear matter. This is because
the contribution from the Landau term becomes less
with inclusion of hyperons due to smaller values of the
coupling strengths of the $\omega$ meson with hyperons 
as compared to coupling strengths with the nucleons. 
The $\rho$ meson mass is observed to drop monotonically 
with density dominantly from the drop in the light quark 
condensate in the medium,
with negligible contribution from the Landau scattering term. 
The effect of isospin 
asymmetry is observed to be small on the masses of the
$\omega$ and $\rho$ mesons, as the dependence on
the light quark condensates is through the combination
$(m_u \bar u u+m_d \bar d d)$, which has marginal
effect from the isospin asymmetry. 
For the $\phi$ meson, there is observed to be
a drop in the mass in nuclear matter due to the modification 
of the strange quark condensate and scalar gluon condensate,
because the contribution from the Landau term for the
$\phi$ meson vanishes in the nuclear matter. The mass shift 
of $\phi$ meson in nuclear medium is seen to be small, of the
order of 20 MeV at a density of 5$\rho_0$. This is because 
the strange condensate as well as gluon condensate 
have very small modification in the medium and occur with
opposite signs in the coefficient ${c^*}_2^\phi$. 
In the presence of hyperons, 
however, there is seen to be an increase in the mass 
of the $\phi$ meson with density due to contribution 
from the Landau term arising from scattering of the
$\phi$ meson with the hyperons. The mass of the $\phi$ meson
is observed to have larger effect from the
Landau scattering term for the isospin symmetric case
as compared to the isospin asymmetric hyperonic matter.

\acknowledgements
The author acknowledges financial support 
from Department of Science and Technology, Government of India 
(project no. SR/S2/HEP-031/2010). 



\begin{thebibliography}{}
\bibitem{rapp} R. Rapp and J. Wambach, Adv. Nucl. Phys. {\bf 25}, 1
(2000).
\bibitem{hatlee}  T. Hatsuda, S.H. Lee, Phys. Rev. C {\bf 46}, (1992) R34.	
\bibitem{hatlee2}  T. Hatsuda, S.H. Lee, H. Shiomi, Phys. Rev. C {\bf 52},
(1995) 3364.	
\bibitem{zschocke} S. Zschocke, O. P. Pavlenko and B. K\"ampfer,
Eur. Phys. Jour. A {\bf 15}, 529 (2002).
\bibitem{klinglnpa}
F.Klingl, N. Kaiser, W. Weise, Nucl. Phys. {\bf A 624},527 (1997).
\bibitem{kwonprc2008}
Y. Kwon, M. Procura and W. Weise, Phys. Rev. {\bf C 78},  055203 (2008).
\bibitem{Abhee}
A.K. Dutt-Mazumder, R. Hofmann, M. Pospelov, Phys. Rev. C {\bf 63}, 
015204 (2000). 
\bibitem{kristof1}
A. Mishra, K. Balazs, D. Zschiesche, S. Schramm,
H. St\"ocker and W. Greiner, Phys. Rev. C {\bf 69}, 024903 (2004).
\bibitem{papa}
 	P. Papazoglou, D. Zschiesche, S. Schramm, J. Schaffner-Bielich,
	H. St\"ocker, and W. Greiner, Phys. Rev. C {\bf 59},  411  (1999).
\bibitem{nstar}
A. Mishra, A. Kumar, S. Sanyal, V. Dexheimer, S. Schramm,
Eur. Phys. Jour. A {\bf  45}, 169 (2010).
\bibitem{vecm}
D. Zschiesche, A. Mishra, S. Schramm, H. St\"ocker and W. Greiner,
Phys. Rev. C  70, 045202 (2004).
\bibitem{isoamss}
A. Mishra, E. L. Bratkovskaya, J. Schaffner-Bielich, S. Schramm
     and H. St\"ocker, Phys. Rev. C {\bf 70}, 044904 (2004).
A. Mishra and S. Schramm, Phys. Rev. C {\bf 74}, 064904 (2006).	
\bibitem{isoamss2}
Amruta Mishra, Arvind Kumar, Sambuddha Sanyal, S. Schramm,
Eur. Phys, J. A {\bf 41}, 205  (2009).  
\bibitem{amarind} 
Amruta Mishra and Arindam Mazumdar, Phys. Rev. C {\bf 79},  024908 (2009). 
\bibitem {amdmeson} 
A. Mishra, E. L. Bratkovskaya, J. Schaffner-Bielich, 
S.Schramm and H. St\"ocker, Phys. Rev. {\bf C 69}, 015202 (2004).
\bibitem{amarvind} 
Arvind Kumar and Amruta Mishra, Phys. Rev. C {\bf 81}, 065204 (2010).
\bibitem{amarvindhyp}
Arvind Kumar and Amruta Mishra, Eur. Phys. J. A {\bf 47}, 164 (2011).
\bibitem{weinberg}
S.Weinberg, Phys. Rev. {\bf 166} 1568 (1968).
\bibitem{coleman}
S. Coleman, J. Wess, B. Zumino, Phys. Rev. {\bf 177} 2239 (1969);
C.G. Callan, S. Coleman, J. Wess, B. Zumino, Phys. Rev. {\bf 177}
2247 (1969).
\bibitem{bardeen}
W. A. Bardeen and B. W. Lee, Phys. Rev. {\bf 177} 2389 (1969).
 \bibitem{sche1}
J. Schechter, Phys. Rev. D {\bf 21}, 3393 (1980).  
\bibitem{ellis}
J.Ellis, Nucl. Phys. {\bf B} 22,  478 (1970);
B. A. Campbell, J. Ellis and K. A. Olive, 
Nucl. Phys. {\bf B} 345, 57 (1990).
\bibitem{heide1}
Erik K. Heide, Serge Rudaz and Paul J. Ellis, Nucl. Phys. A {\bf 571}, (2001) 713.
\bibitem{chqsram}
Arvind Kumar and Amruta Mishra, Phys. Rev. C {\bf 82}, 045207 (2010). 
\bibitem{cohen}
Thomas D. Cohen, R. J. Furnstahl and David K. Griegel,
 Phys. Rev. C {\bf 45}, 1881 (1992).
\bibitem{svznpb1}
M. A. Shifman, A. I. Vainshtein and V. I. Zakharov, Nucl. Phys. {\bf Bb 147},
385 (1979).
\bibitem{svznpb2}
M. A. Shifman, A. I. Vainshtein and V. I. Zakharov, Nucl. Phys. {\bf Bb 147},
448 (1979).
\bibitem{bochkarev}
A. I. Bochkarev and M. E. Shaposhnikov, Phys. Lett. B {\bf 145}, 276 (1984);
ibid, Nucl. Phys. B {\bf 268}, 220 (1986).
\bibitem{florkowski}
W. Florkowski, W. Broniowski, Nucl. Phys. A {\bf 651}, 397 (1999).

\end{thebibliography}
\end{document}